\documentclass[sigconf]{acmart}

\usepackage[T1]{fontenc}
\usepackage{kantlipsum}
\usepackage[breakable, theorems, skins]{tcolorbox}
\usepackage{hyperref}
\tcbset{enhanced}

\DeclareRobustCommand{\prompt}[2][gray!10]{%
\begin{tcolorbox}[   
        breakable,
        left=2pt,
        right=2pt,
        colback=#1,
        colframe=#1,
        width=\dimexpr\columnwidth\relax, 
        enlarge left by=0mm,
        boxsep=0pt,
        arc=0pt,outer arc=0pt,
        ]
        #2
\end{tcolorbox}
}

\DeclareRobustCommand{\response}[2][gray!30]{%
\begin{tcolorbox}[   
        breakable,
        left=2pt,
        right=2pt,
        colback=#1,
        colframe=#1,
        width=\dimexpr\columnwidth\relax, 
        enlarge left by=0mm,
        boxsep=0pt,
        arc=0pt,outer arc=0pt,
        ]
        #2
\end{tcolorbox}
}

\AtBeginDocument{%
  }

\acmConference[ASPLOS `23 WACI]{Wild and Crazy Ideas, ASPLOS 2023}{March 26th, 2023}{Vancouver, Canada}

\setcopyright{none}
\renewcommand\footnotetextcopyrightpermission[1]{} 

\newcommand{\llamas}{LLaMaS}

\newcommand{\ignore}[1]{}
\newcommand{\ul}[1]{\underline{#1}}

\begin{document}

\title{Herding LLaMaS: Using LLMs as an OS Module}

\author{Aditya K Kamath}
\authornote{Both authors contributed equally to this research.}
\affiliation{%
  \institution{University of Washington}
  \city{Seattle}
  \state{Washington}
  \country{USA}
}
\email{akkamath@uw.edu}

\author{Sujay Yadalam}
\authornotemark[1]
\affiliation{%
  \institution{University of Wisconsin--Madison}
  \city{Madison}
  \state{Wisconsin}
  \country{USA}
}
\email{sujayyadalam@cs.wisc.edu}

\renewcommand{\shortauthors}{A.K. Kamath and S. Yadalam}



\maketitle

\section{Introduction}
\label{sec:Introduction}

Computer systems are becoming increasingly heterogeneous with the emergence of new memory technologies and compute devices. 
GPUs alongside CPUs have become commonplace and CXL is poised to be a mainstay of cloud systems.
The operating system is responsible for managing these hardware resources, requiring modification every time a new device is released. 
Years of research and development are sunk into tuning the OS for high performance with each new heterogeneous device~\cite{Achermann:ASPLOS:2020, Yan:ASPLOS:2019, Rayback:SOSP:2021, Dashti:ASPLOS:2013, Huaicheng:ASPLOS:2023, Maruf:ASPLOS:2023, Agarwal:ASPLOS:2015, Ausavarungnirun:MICRO:2017, Panwar:ASPLOS:2021}.
With the recent explosion in memory technologies and domain-specific accelerators, it would be beneficial to have an OS that could provide high performance for new devices without significant effort.

We propose \llamas{} which can adapt to new devices easily. \llamas{} uses Large Language Models (LLMs) to extract the useful features of new devices from their textual description and uses these features to make operating system decisions at runtime. Adding support to \llamas{} for a new device is as simple as describing the system and new device properties in plaintext.



\llamas{} reduces the burden on system administrators to enable easy integration of new devices into production systems.

Preliminary evaluation using ChatGPT~\cite{ChatGPT} shows that LLMs are capable of extracting device features from text and make correct OS decisions based on those features.

\section{Burden of heterogeneity on the OS}
\ignore{
\begin{enumerate}
    \item Many advances in technologies. Our focus is on two, memory and compute. Memory: NUMA, NVM, CXL, HBM. Compute: GPUs, accelerators, FPGA, TPUs.
    \item OS is responsible for the interface between the app and devices. As such, every new tech requires OS changes for support.
    \item Significant research effort spent on improving performance. List papers.
\end{enumerate}
}

The end of Moore's law and Dennard scaling has made the use of heterogeneous systems necessary. 
Modern high-performance systems are embracing heterogeneity in both memory and compute. 
These systems combine the best properties of different memory technologies to optimize for latency, bandwidth, capacity, and cost. 
For processors, the adoption of GPUs and other domain-specific accelerators (DSAs) has helped push the boundaries of compute.

Different applications exhibit different memory requirements, necessitating a diverse set of memory devices to satisfy all of them. 
A modern HPC system could be connected to local DRAM and NVM, and have disaggregated memory over CXL. 
Non-volatile memory (NVM)~\cite{Optane} provides high capacities, but experiences read/write asymmetry as well as reduced bandwidth. 
Similarly, CXL provides greater memory capacity than on-board DRAM, at the expense of increased latencies and lower bandwidth~\cite{Huaicheng:ASPLOS:2023}.

Data-intensive applications like machine learning or scientific computations require high throughput that is not met by conventional architectures. 
This has led to the development of accelerators such as GPUs and specialized hardware.
In the face of this explosive growth of diverse DSAs each with their own unique API, there has been significant effort being put into unifying application development~\cite{ASPLOS21_Keynote}. 
The RISC-V group endeavors to provide a unified ISA that can support the unique attributes that these different accelerators require~\cite{SiFive}.
On the compiler side, the MLIR project~\cite{Lattner:CGO:2021} is providing an intermediate layer that allows developers to code in their language of choice and then compile the source code into optimized binaries for a chosen processor. 
In the face of these advancements, we envision a future where an application binary could be deployed on any processing device without programmer input. 
The operating system (OS) would be tasked with selecting the optimal processing device for the application.

The operating system is the gatekeeper between applications and hardware devices.
Beyond providing minimal support for these devices, the OS must be aware of the different intricacies and characteristics under which the devices perform optimally, to remove the reliance on application programmers.

This requirement of OS modification leads to a significant amount of research effort being spent in trying to devise the best method for handling these devices. 
For example, there has been significant work in page placement for NVM~\cite{Yan:ASPLOS:2019, Rayback:SOSP:2021} and CXL~\cite{Huaicheng:ASPLOS:2023, Maruf:ASPLOS:2023}.
In addition, many works have explored techniques for managing data placement and replication for NUMA systems~\cite{Dashti:ASPLOS:2013, Achermann:ASPLOS:2020}. Similarly, we foresee that significant effort will need to be made in order to allow the OS to select the optimal processing device.

%


It would be beneficial to have an operating system that could adapt to any heterogeneous system quickly. 
Such an operating system would reduce the burden of researchers and system administrators. 
It would also reduce the effort required to integrate new devices into production systems.

\section{Our system: Herding LLaMaS}
Our goal is to design an operating system that would be able to (1) adapt to new heterogeneous technologies while (2) requiring minimal intervention from the programmer/system administrator.

To this end, we propose using \ul{L}arge \ul{La}nguage \ul{M}odels \ul{a}s an O\ul{S} Module (\llamas{}) for managing \ul{he}terogeneous \ul{r}esources and \ul{d}evices (a.k.a., the herd)\footnote{It is worth noting that for our title "Herding \llamas{}", \llamas{} is responsible for managing the herd, and so is performing the herding, not being herded.}.

Language models are a class of Natural Language Processing algorithms that aim to recognize, summarize, translate, predict and generate text. 
Large language models or LLMs are models trained on very large datasets with billions to trillions of parameters. 
OpenAI's GPT-3 has over 100 billion parameters and is trained on almost the entirety of the internet. 
The recent success of LLMs is attributed to few-shot or zero-shot learning. 
LLMs can solve various tasks by simply training the models on a few examples (few-shot) or by providing instructions describing the task (zero-shot).

\llamas{} takes advantage of LLMs' ability to perform zero-shot learning. \llamas{} is able to flexibly adapt to new devices with quirky characteristics, requiring only a simple description of the system and hardware characteristics.

\begin{figure}
    \setlength{\belowcaptionskip}{-20pt}
    \centering
    \includegraphics[width=\linewidth]{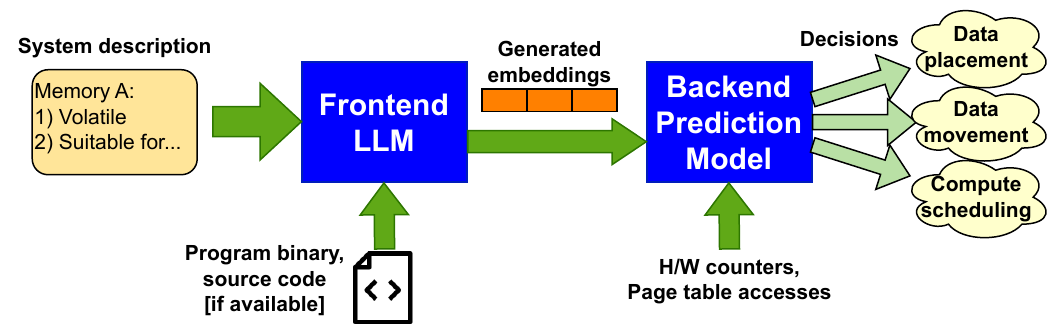}
    \caption{LLaMaS system design.}
    \label{fig:sys_design}
\end{figure}

Figure \ref{fig:sys_design} describes how \llamas{} functions.
There are two main components of \llamas{}: (1) frontend LLM and (2) backend prediction model.
The frontend LLM takes the system description as a text file along with the program binary and/or source code if available.
It processes them to identify the important features and patterns that could influence decisions during program execution and generates embeddings that contain these information.

The backend prediction model (BPM) is responsible for making runtime OS decisions. The embeddings generated by the frontend LLM guide the BPM on  which heuristics influence decisions and when the decisions need to be made. For instance, the LLM could inform the BPM to migrate data when the accesses per second to a page exceed a certain value.


As an example, the frontend LLM may be informed that a system contains two types of memory, repetitive access tech (RAT) which provides low latency for frequently accessed data but high latency for infrequently accessed data, and DRAM which provides predictable latencies for all data.
Through the embeddings, it will inform the BPM that frequently accessed data should be kept in RAT.
During an application's runtime, the hardware counters may indicate that data in DRAM is being frequently accessed, causing the BPM to move it to RAT.
Through this process, the OS teaches itself how to handle new technology with minimal external support.


\label{sec:design}

\ignore{
\begin{enumerate}
    \item What we require from our system. Reiteration of problems and goals.
    \item Our proposal: Herding LLaMaS, and what it hopes to achieve.
    \item In-depth description of Figure \ref{fig:sys_design}.
\end{enumerate}
}

\section{Feasibility study}
\label{sec:feasibility_study}

To study the feasibility of \llamas, we conducted experiments using ChatGPT~\cite{ChatGPT}, a chatbot built on top of OpenAI's GPT-3.5 and GPT-4 language models. 
The goal of the experiments was to understand the abilities of a language model to extract insights from a textual description of device characteristics. 
The prompts to ChatGPT had 2 components: (1) descrition of a heterogeneous system configuration and device properties, and (2) OS decisions to be made. 
Below we share the results of 3 such experiments. 
We first show the prompt to ChatGPT then its response:

\textbf{Experiment 1: } Here, we wanted to ensure that LLMs could understand system descriptions to extract useful features.
\prompt{\textbf{Prompt}: "System you manage has CPU and GPU. CPU is connected to DRAM and GPU is connected to HBM. For the GPU matrix multiplication program shown below, identify variables that should be allocated on DRAM and those on HBM."}
\vspace{-15pt}
\response{\textbf{ChatGPT}: "Variable h\_A is used to store the input matrix on the host, so allocate in host memory. Variable d\_A is used to store the input matrix on the device, so allocate in device memory."}

ChatGPT was able to successfully understand the system description. Since ChatGPT had been trained on data involving CPUs and GPUs, we didn't have to describe the system to it. 
It was able to understand the use of different variables and map them to the host or device memory.\\

\textbf{Experiment 2: } While the above experiment proved that LLMs could work well on data it was trained on, \llamas{} requires that the OS module adapts to new devices with unseen characteristics. Therefore, our second experiment imagined a heterogenous system with a futuristic device.

\prompt{\textbf{Prompt}: "System you manage has CPU connected to DRAM and Repetitive Access Technology (RAT) memories. RAT is a new memory technology that provides low latency for frequently accessed data. Given below program, identify variables that should be allocated in RAT memory.}
\vspace{-15pt}
\response{\textbf{ChatGPT}: "RAT memory should be used for variables that are accessed repeatedly such as arrays or matrices. DRAM should be used for variables that are accessed randomly and infrequently. These variables should be allocated in RAT: priceArr, avg, ..."}

ChatGPT was able to conceptualize the new memory technology and capture its properties. It identified the important features to consider while making memory placement decisions.\\

\textbf{Experiment 3: Task placement:} Above experiments demonstrated that abilities of LLMs to aid in memory management. We then tried to understand the abilities of LLMs to perform task mapping to different computational platforms.

\prompt{\textbf{Prompt}: You manage a system with a CPU of 24 cores and a GPU. For best performance, where would you run the following functions?}
\vspace{-10pt}
\response{\textbf{ChatGPT}: Func\_A involves matrix multiplication and is highly parallelizable. Therefore, we would prefer to run it on the GPU. On the other hand, Func\_B involves pointer manipulation and memory operations, so we would prefer to run on the CPU.}

ChatGPT was able to understand functions' goals and properties (parallelizable, memory access patterns) and match it with the properties of the underlying hardware.


\bibliographystyle{ACM-Reference-Format}
\bibliography{text/references}

\appendix

\end{document}